\documentclass{jpp}
\usepackage{graphicx}

\usepackage[utf8]{inputenc}
\usepackage[T1]{fontenc}
\usepackage{amsmath}

\usepackage{mathdots}
\usepackage{graphicx}
\usepackage{nccmath}
\usepackage{bbold}
\usepackage{amsfonts}
\usepackage{xspace}
\usepackage{xcolor}
\usepackage{upgreek}
\usepackage{silence}

\newcommand{\ie}{i.e.\@\xspace}

\newcommand{\m}[1]{$\smash{#1}$}

\renewcommand{\eq}[1]{(\ref{#1})}
\newcommand{\Eq}[1]{\eq{#1}}
\newcommand{\Eqs}[1]{\eq{#1}}
\newcommand{\Fig}[1]{figure~\ref{#1}}
\newcommand{\Figs}[1]{figures~\ref{#1}}
\newcommand{\Sec}[1]{section~\ref{#1}}
\newcommand{\Secs}[1]{sections~\ref{#1}}
\newcommand{\App}[1]{appendix~\ref{#1}}

\newcommand{\mc}[1]{\mathcal{#1}}

\renewcommand{\vec}[1]{\boldsymbol{#1}}
\newcommand{\uvec}[1]{\vec{e}_{#1}}
\newcommand{\matr}[1]{\vec{#1}}
\newcommand{\oper}[1]{\hat{#1}}
\newcommand{\boper}[1]{\hat{\vec{#1}}}

\newcommand{\ii}{\mathrm{i}}
\newcommand{\dd}{\mathrm{d}}
\newcommand{\ee}{\mathrm{e}}
\newcommand{\pd}{\partial}
\newcommand{\del}{\nabla}
\newcommand{\Obig}{\mc{O}}

\newcommand{\const}{\text{const}}

\newcommand{\amp}{\psi}

\DeclareMathOperator{\re}{Re}
\DeclareMathOperator{\im}{Im}
\DeclareMathOperator{\sgn}{sgn}

\newcommand{\overbar}[1]{\mkern 1.5mu\overline{\mkern-1.5mu#1\mkern-1.5mu}\mkern 1.5mu}

\definecolor{darkred}{rgb}{0.75, 0, 0}
\usepackage[colorlinks=true, linkcolor=darkred, citecolor=darkred, hypertexnames=false]{hyperref}

\newenvironment{mpmatrix}%
 {\begin{medsize}\begin{pmatrix}}%
 {\end{pmatrix}\end{medsize}}

\shorttitle{Modulational dynamics in MHD}
\shortauthor{S. Jin and I. Y. Dodin}

\title{On reduced modeling of the modulational dynamics in magnetohydrodynamics}

\author{S. Jin\aff{1, 2}
  \corresp{\email{sjin@pppl.gov}} \and
  I. Y. Dodin\aff{1, 2}}

\affiliation{\aff{1}Department of Astrophysical Sciences, Princeton University, Princeton, New Jersey 08543, USA
\aff{2}Princeton Plasma Physics Laboratory, Princeton, New Jersey 08540, USA}

\begin{document}

\maketitle

\begin{abstract}
This paper explores structure formation in two-dimensional magnetohydrodynamic (MHD) turbulence as a modulational instability (MI) of turbulent fluctuations. We focus on the early stages of structure formation and consider simple backgrounds that allow for a tractable model of the MI while retaining the full chain of modulational harmonics. This approach allows us to systematically examine the validity of popular closures such as the quasilinear approximation and other low-order truncations. We find that, although such simple closures can provide quantitatively accurate approximations of the MI growth rates in some regimes, they can fail to capture the modulational dynamics in adjacent regimes even qualitatively, falsely predicting MI when the system is actually stable. We find that this discrepancy is due to the excitation of propagating spectral waves (PSWs) which can ballistically transport energy along the modulational spectrum, unimpeded until dissipative scales, thereby breaking the feedback loops that would otherwise sustain MIs. PSWs can be self-maintained as global modes with real frequencies and drain energy from the primary structure at a constant rate until the primary structure is depleted. To describe these waves within a reduced model, we propose an approximate spectral closure that captures them and MIs on the same footing. We also find that introducing corrections to ideal MHD, conservative or dissipative, can suppress PSWs and reinstate the accuracy of the quasilinear approximation. In this sense, ideal MHD is a `singular' system that is particularly sensitive to the accuracy of the closure within mean-field models.
\end{abstract}

\section{Introduction}
\label{sec:intro}

\subsection{Motivation}

Coherent-structure formation from turbulence is ubiquitous in nature, intrinsically compelling, and one of the precious few aspects of turbulence that are relatively yielding to analytical efforts \citep{ref:hussain86, ref:karimabadi13, ref:smolyakov00, ref:krashennikov08}. For magnetohydrodynamic (MHD) turbulence \citep{book:biskamp03, ref:beresnyak19, ref:schekochihin22} and the turbulent-dynamo problem \citep{ref:brandenburg12, ref:tobias21, ref:rincon16}, this has spawned a vast body of work known as mean-field electrodynamics, or mean-field dynamo theory \citep{ref:radler07, book:krause80, book:moffatt78, ref:brandenburg18, ref:gruzinov94}. In such approaches, the velocity and magnetic fields are split into fluctuations and mean, or average, fields (the definition of average depends on the problem); then, various closures are implemented to obtain the mean-field evolution in response to the fluctuations \citep{ref:blackman02, ref:brandenburg05, ref:pouquet76, ref:kraichnan77, ref:nicklaus88}.

These closures ultimately rely on a truncation of a formally infinite chain of correlation functions of increasingly high order.\footnote{Here, we use the term `truncation' to refer to low-order closures in general, as opposed to precisely the closure choice of setting all correlations to zero after some cutoff. The latter, which we will refer to as a simple truncation, is done in the quasilinear approximation discussed below but not, for example, in the various \m{\tau} approximations \citep{ref:brandenburg05, ref:blackman02}.} This is a common approach to reduced modeling of nonlinear systems, and one can expect it to work when the fluctuations are small compared to the mean field. However, in turbulent structure formation, the usual roles of perturbation and background are switched, that is, mean fields emerge as small modulations of order-one correlation functions that characterize turbulent fluctuations. Hence, the question of the validity of low-order closures becomes nontrivial.

In this paper, we are concerned with the popular closure called the first-order smoothing \citep{book:krause80} or second-order correlation approximation \citep{ref:radler82}. It implies that the mean-field evolution is determined only by second-order correlations of the fluctuations, which, in a broader context, is also known as the quasilinear approximation \citep{my:ql}. On the one hand, one can expect its validity domain to be extremely limited. It can be rigorously justified only for small magnetic Reynolds numbers or short correlation times \citep{ref:brandenburg05_rev, ref:rincon19}, and comparisons with direct numerical simulations often corroborate this expectation \citep{ref:pipin08, ref:schrinner05}. On the other hand, the quasilinear approximation endures in its popularity as a workhorse that is too convenient to cast aside in analytical calculations \citep{ref:squire15_a, ref:lingam16, ref:masada14, ref:gopalakrishnan23}. Furthermore, in some cases, quasilinear calculations have been found to produce good agreement with direct numerical simulations (DNS) even outside of their formal validity domain \citep{ref:kapyla06}.\footnote{Confusingly, various analytical closures and DNS can conflict and agree in various combinations \citep{ref:zhou21}. But this issue is beyond the scope of the present paper.} It is therefore important to understand when, and how exactly, the quasilinear approximation fails. Answering this question requires exploring a mean-field model that retains high-order correlations. That is what the present paper is about.

\subsection{Outline}

Here, we explore structure formation in MHD turbulence as a modulational instability \citep{ref:zakharov09} (MI) of the flow-velocity and magnetic-field fluctuations comprising the background turbulence. (For a recent overview of this approach in application to drift-wave and hydrodynamic turbulence, see \citet{my:wkeadv, my:navier}.) For simplicity, we restrict our consideration to two-dimensional (2-D) turbulence. Although bounded planar flows cannot sustain long-lasting magnetic fields in two dimensions (an anti-dynamo theorem by \citet{ref:zeldovich57}), 2-D turbulence nevertheless hosts rich physics and has long served as popular testing grounds for various aspects of MHD turbulence theory due to its (relative) tractability \citep{ref:pouquet78, ref:biskamp01, ref:nazarenko07, ref:degiorgio17, ref:biskamp96}. The goal of this paper is to complement those works with a detailed study of MIs using both analytical calculations and DNS.

\begin{figure*}
 \centering
 \includegraphics{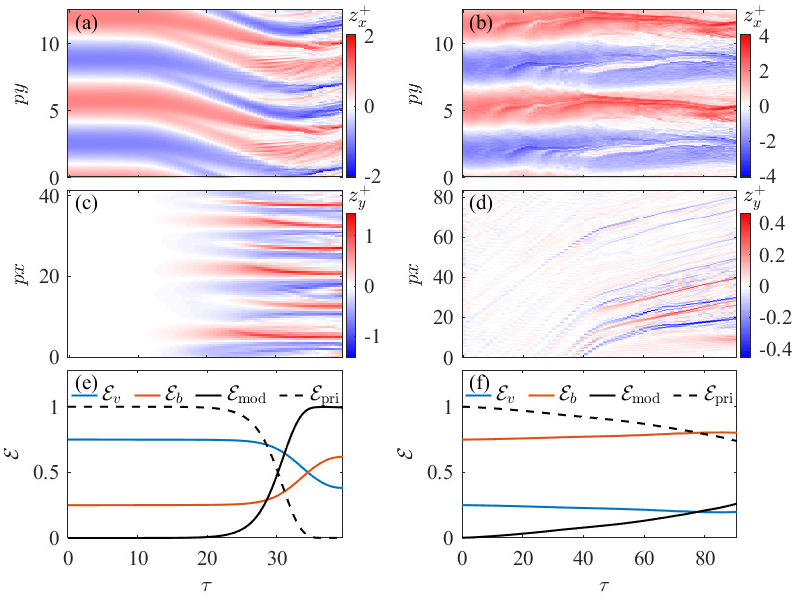}
 \caption{\label{fig:spaceplot}
 DNS of modulational dynamics seeded at \m{\tau=0} with random noise. (This figure is discussed in detail in later sections. For the notation, see \Secs{sec:base} and \ref{sec:primarymodecoupling}.) Figures (a), (c), and (e) correspond to a modulationally unstable primary mode \eq{eq:prifields} with \m{\theta^\pm=\pm\upi/6}, and (b), (d), and (f) correspond to a modulationally stable primary mode with \m{\theta^\pm=\pm\upi/3}. Figures~(a) and~(b) show \m{z_x^+(t,y)}, (c)~and (d)~show \m{z_y^+(t,x)}, and~(e) and~(f) show the corresponding energy breakdown. Specifically, \m{\mc{E}_v} is the normalised kinetic energy \eq{eq:normEv}, \m{\mc{E}_b = E_b/E_{\text{tot}}} is the normalised magnetic energy \eq{eq:normEb}, \m{\mc{E}_{\text{mod}}} is the normalised modulational-mode energy \eq{eq:normEmod}, and \m{\mc{E}_{\text{pri}}} is the normalised primary-mode energy \eq{eq:normEpri}. The colour bar shows the field amplitudes normalised to \m{\mc{A}/p}. The spatial coordinates are normalised to \m{1/p}.}
\end{figure*}

Specifically, we study how energy is transferred from primary fluctuations to mean magnetic field and mean velocity during the early stages of structure formation, when these mean fields are weak and the primary structure can be treated as fixed. To analyze this process in detail, we assume a maximally reduced model of the turbulent fluctuations, namely, a spatially monochromatic primary structure. The simplicity of this model has been detracting attention from it in the literature in favor of more realistic turbulence settings, but there is more to it that meets the eye. Even with a monochromatic primary structure, the modulational dynamics remains remarkably intricate and, depending on the system parameters, can change drastically (\Fig{fig:spaceplot}) in ways that are not captured by the common closures or obvious from the standard eigenmode analysis \citep{book:friedberg}. However, the simplicity of our model helps unravel this mystery.

We find that, when the primary structure \textit{truly} experiences a MI, this MI can be described well with typical truncated models that neglect high-order correlations. However, already in adjacent regimes, such truncated models can fail spectacularly and produce false positives for instability. To study this process in detail, we propose an `extended' quasilinear theory (XQLT) that treats the primary structure as fixed but includes the entire spectrum of modulational harmonics (as opposed to just the low-order harmonics, as usual). From XQLT, also corroborated by DNS, we find that the difference between said regimes is due to a fundamental difference in the structure of the modulational spectrum. For unstable modes, the spectrum is exponentially localised at low harmonic numbers, so truncated models are justified. But this localization does not always occur. At other parameters, modulational modes turn into constant-amplitude waves propagating down the spectrum, unimpeded until dissipative scales. These spectral waves are self-maintained as global modes with real frequencies and cause ballistic energy transport along the spectrum, breaking the feedback loops that could otherwise sustain MI. The ballistic transport drains energy from the primary structure at a constant rate until the primary structure is depleted. Notably, these effects are overlooked if one assumes from the start that the modulation spectrum is confined and that, accordingly, dissipation entirely vanishes in the limit of zero viscosity and resistivity.

A comment is due here regarding how the established truncated MHD closures fit into this picture. Given the indispensability of such simple closures, a higher-resolution understanding of their limitations can help guide and interpret their inevitable application. For example, we show that both conservative and dissipative corrections to ideal MHD tend to restore the validity of the quasilinear approximation. In this sense, ideal incompressible MHD is a `singular' theory that is particularly sensitive to the accuracy of the closure within mean-field models. That said, it is an important conclusion of our work that, unless deviations from ideal incompressible MHD are substantial, changing the turbulence parameters even slightly can produce qualitative inaccuracies in an otherwise workable reduced model. Therefore, efforts to extrapolate closures or infer them from general considerations should be supplemented with first-principle calculations based on the underlying modulational dynamics. As a step in this direction, we propose a spectral closure that captures both the propagating spectral waves and~MIs within our model. (But, of course, more work remains to be done for generic MHD turbulence.)

The paper is organised as follows. In \Sec{sec:XQLT}, we present the basic equations and derive the XQLT. In \Sec{sec:trunc}, we examine the ability of truncated models of low-order to predict the MI rate. In \Sec{sec:spectralw}, we introduce spectral waves and describe their properties. In \Sec{sec:closure}, we propose a closure that can simultaneously capture MIs and spectral waves. In \Sec{sec:dispersion}, we demonstrate the impact of corrections to ideal MHD on the modulational dynamics. In \Sec{sec:summary}, we summarise our main results. Auxiliary calculations are presented in appendices. 

\section{Extended quasilinear theory}\label{sec:XQLT}

\subsection{Base model}
\label{sec:base}

\subsubsection{Incompressible resistive MHD}

As a base model, we assume incompressible resistive MHD with homogeneous mass density \m{\rho = \const}. The corresponding governing equations are
\begin{subequations}\label{eq:mhd}
\begin{gather}
 \pd_t \vec{v} + (\vec{v} \cdot \nabla) \vec{v} 
  = (\vec{b} \cdot \nabla)\vec{b} - \nabla P 
    + \nu \del^2 \vec{v},
    \label{eq:v}\\
 \pd_t \vec{b} = \nabla \times (\vec{v} \times \vec{b}) + \eta \nabla^2 \vec{b}.
    \label{eq:b}
\end{gather}
Here \m{\vec{v}} is the fluid velocity; \m{\vec{b} \doteq \vec{B}/\sqrt{4\upi\rho}} is the local Alfv\'en velocity 
(the symbol \m{\doteq} denotes definitions), \ie rescaled magnetic field \m{\vec{B}}; and \m{P \doteq (P_{\text{kin}} + B^2/8\upi)/\rho} is the normalised total pressure, with \m{P_{\text{kin}}} being the kinetic pressure. Also, \m{\nu} is the viscosity  and \m{\eta} is the resistivity, both of which are assumed either constant or negligible (except at small enough scales). Due to incompressibility and magnetic Gauss's law, one also has
\begin{gather}
 \nabla \cdot \vec{v} = 0,
 \qquad
 \nabla \cdot \vec{b} = 0.
\end{gather}
\end{subequations}

In order to put \Eqs{eq:mhd} in a more symmetric form, let us rewrite them in terms of the two Els\"asser fields \m{\vec{z}^\pm} \citep{ref:elsasser50}, which are also solenoidal:
\begin{gather}\label{eq:zdef}
 \vec{z}^\pm \doteq \vec{v} \pm \vec{b},
 \qquad
 \del \cdot \vec{z}^\pm  = 0.
\end{gather}
This leads to two coupled equations for \m{\vec{z}^\pm}:
\begin{gather}\label{eq:elsasser}
\pd_t \vec{z}^\pm=-(\vec{z}^\mp\cdot\nabla)\vec{z}^\pm-\nabla P
+\nu_+ \nabla^2 \vec{z}^\pm+\nu_- \nabla^2 \vec{z}^\mp,
\end{gather}
where \m{\nu_+ \doteq (\nu + \eta)/2} and \m{\nu_- \doteq (\nu - \eta)/2}. The normalised total pressure \m{P} can be found as follows. By taking the divergence of \Eq{eq:v} and using \Eq{eq:zdef}, one obtains
\begin{gather}\label{eq:peq}
 \nabla^2 P = -\nabla\cdot[(\vec{z}^\mp \cdot \nabla)\vec{z}^\pm].
\end{gather}
Let us introduce the wavevector operator \m{\boper{k} \doteq - \ii \nabla}, so \m{\oper{k}^2 \doteq \smash{\boper{k}}^2 = - \nabla^2}. Let us also assume some appropriate (say, periodic) boundary conditions. Then, \Eq{eq:peq} yields
\begin{gather}
 P = -\oper{k}^{-2}\boper{k}\cdot[(\vec{z}^\mp \cdot \boper{k})\vec{z}^\pm] + \const,
\end{gather}
whence \m{\nabla P} in \Eq{eq:elsasser} can be expressed through \m{\vec{z}^\pm}. Alternatively, \m{\nabla P} can be eliminated from \Eq{eq:elsasser} by taking the curl of this equation and rewriting the result in terms of the Els\"asser vorticities
\begin{gather}
\vec{w}^\pm \doteq \nabla \times \vec{z}^\pm.
\end{gather}
Indeed, due to \Eq{eq:zdef}, one can express \m{\vec{z}^\pm} using a vector potential \m{\vec{a}^\pm} such that \m{\vec{z}^\pm = \nabla \times \vec{a}^\pm}. Let us assume the gauge such that \m{\nabla \cdot \vec{a}^\pm = 0}. Then,
\begin{gather}
 \vec{w}^\pm = \nabla \times (\nabla \times \vec{a}^\pm)
 = - \nabla^2 \vec{a}^\pm \equiv \oper{k}^2 \vec{a}^\pm,
\end{gather}
whence
\begin{gather}
 \vec{z}^\pm = \ii \oper{k}^{-2} (\boper{k} \times \vec{w}^\pm).
\end{gather}
(Here, we assume that \m{\vec{z}^\pm} have zero spatial average, so \m{\oper{k}^2} is invertible.) Then, \Eq{eq:elsasser} can be expressed through \m{\vec{w}^\pm} alone:
\begin{gather}
\pd_t \vec{w}^\pm = 
-\boper{k} \times
\{
[(\boper{k} \times \oper{k}^{-2}\vec{w}^\mp)\cdot \boper{k}]
(\boper{k}\times\oper{k}^{-2}\vec{w}^\pm)
\}
-\oper{k}^2(\nu_+ \vec{w}^\pm + \nu_- \vec{w}^\mp).
\label{eq:vecw}
\end{gather}

\subsubsection{2-D collisionless model}

For simplicity, below we adopt a 2-D model, meaning that \m{\vec{z}^\pm} lie in the \m{(x, y)} plane and \m{\pd_z = 0}. In this case, the only potentially nonzero component of \m{\vec{w}^\pm} is the \m{z}-component, \m{w^\pm \doteq (\nabla\times\vec{z}^\pm)_z}. We also forgo an explicit treatment of the viscosities \m{\nu_\pm} below. (However small, though, these viscosities remain nonnegligible in that they determine absorbing boundary conditions at infinity in the spectral space; see \Sec{sec:spectralw}.) Then, the vector equation \eq{eq:vecw} can be replaced with a scalar equation for \m{w^\pm}: 
\begin{gather}
\label{eq:scalarw}
\pd_t w^\pm = \epsilon_{lmz} \Bigg[
\left(\frac{\oper{k}_m}{\oper{k}^2}\, w^\mp \right)\left(\oper{k}_l w^\pm \right)
- \left(\frac{\oper{k}_n\oper{k}_l}{\oper{k}^2}\, w^\mp\right) 
\left(\frac{\oper{k}_n\oper{k}_m}{\oper{k}^2} \, w^\pm\right)
\Bigg],
\end{gather}
where \m{\epsilon_{lmz}} is the Levi--Civita symbol with the third index fixed to \m{z}.  Note that the right-hand side of \Eq{eq:scalarw} vanishes whenever \m{w^+} or \m{w^-} is zero. This means that any stationary \m{w^+} is a solution as long as \m{w^-} is zero and vice versa.

\subsubsection{Fourier representation}
\label{sec:fourier}

Assuming the Fourier representation
\begin{gather}
w^\pm = \sum_{\vec{k}} w_{\vec{k}}^\pm(t) \ee^{\ii \vec{k} \cdot \vec{x}},
\end{gather}
where \m{\vec{x} \doteq (x, y)}, one arrives at the following equation for the Fourier coefficients \m{w_{\vec{k}}^\pm}:
 \begin{gather}\label{eq:fourier}
 \pd_t w^\pm_{\vec{k}}
  = \sum_{\vec{k}_1,\vec{k}_2} T(\vec{k}_1,\vec{k}_2)
    w^\mp_{\vec{k}_1} w^\pm_{\vec{k}_2} 
    \delta_{\vec{k},\vec{k}_2+\vec{k}_1},
\end{gather}
where the dot is the time derivative, \m{\delta} is the Kronecker delta, \m{T(\vec{k}_1,\vec{k}_2)} are the coupling coefficients given by
\begin{gather}\label{eq:T}
 T(\vec{k}_1,\vec{k}_2)
  = [\uvec{z} \cdot (\vec{k}_2\times\vec{k}_1)]\,
    \frac{(\vec{k}_2 + \vec{k}_1)\cdot\vec{k}_2}{k_1^2 k_2^2},
\end{gather}
and \m{\uvec{j}} is the unit vector along the \m{j}th axis. Accordingly, the kinetic and magnetic energy are given by
\begin{subequations}
 \begin{align}
 E_v 
  & = \frac{\rho}{8}\,\sum_{\vec{k}}
    \frac{1}{k^2}\,|w_{\vec{k}}^+ + w_{\vec{k}}^-|^2
    \equiv \sum_{\vec{k}} E_{{v},\vec{k}},\\
 E_b
  & = \frac{\rho}{8}\,\sum_{\vec{k}}\frac{1}{k^2}\,
    |w_{\vec{k}}^+ - w_{\vec{k}}^-|^2
  \equiv \sum_{\vec{k}} E_{{b},\vec{k}},
 \end{align}
\end{subequations}
where \m{E_{{v},\vec{k}}}, and \m{E_{{b},\vec{k}}} are the corresponding spectral energy densities. Then, the total energy density is
\begin{gather}
 E = \frac{\rho}{4}\,\sum_{\vec{k}}
     \frac{1}{k^2}\,
     (|w_{\vec{k}}^+|^2 + |w_{\vec{k}}^-|^2)
\equiv \sum_{\vec{k}} E_{\vec{k}},
\end{gather}
where \m{E_{\vec{k}}} is the spectral total-energy density.

\subsection{Harmonic coupling mediated by a primary structure}
\label{sec:primarymodecoupling}

\subsubsection{Equation for the modulational spectrum}
\label{sec:firstorder}

Let us explore modulational stability of a Fourier harmonic with a given wavevector \m{\vec{k} = \vec{p}} and the corresponding order-one Fourier amplitudes \m{w_{\vec{p}}^\pm}. We will call it the primary harmonic or, if multiple \m{\vec{p}}s are present, primary structure. Let us choose axes such that \m{\vec{p} = p\uvec{y}} and assume a perturbation with a wavevector \m{\vec{q}}. As can be seen from \Eq{eq:T}, nontrivial coupling requires a component of \m{\vec{q}} perpendicular to \m{\vec{p}}, and the component of \m{\vec{q}} parallel to \m{\vec{p}} does not qualitatively affect the modulational dynamics.\footnote{As shown below, the modulational dynamics is determined mostly by the asymptotic form of the coefficients \m{\alpha_n^\pm} and \m{\beta_n^\pm} (introduced in \Sec{sec:dimeq}) at large \m{n}, and this form is insensitive to the component of \m{\vec{q}} parallel to \m{\vec{p}}.} For simplicity then, let us assume \m{\vec{q} = q\uvec{x}}.  Let us also assume \m{w_{\vec{p}}^\pm = \Obig(1)}, \m{w_{\vec{q}+n\vec{p}}^\pm = \Obig(\epsilon)}, and same for any other \m{w_{\vec{k} \ne \vec{p}}^\pm}, where \m{\epsilon \ll 1} is a small parameter. (Remember that \m{\Obig(\epsilon)} means, loosely speaking, `scales as \m{\epsilon} or smaller' but not necessarily `of order \m{\epsilon}'.) Then, by linearizing \Eq{eq:fourier} in \m{\epsilon}, one obtains the following chain of equations:
\begin{align}
\pd_t w^\pm_{\vec{k}}
 \approx  \,\,&T(\vec{p},\vec{k}_-)w^\mp_{\vec{p}} w^\pm_{\vec{k}_-} + T(-\vec{p},\vec{k}_+)w^\mp_{-\vec{p}} w^\pm_{\vec{k}_+}
  \notag\\
  + & T(\vec{k}_-,\vec{p})w^\mp_{\vec{k}_-} w^\pm_{\vec{p}} 
+ T(\vec{k}_+,-\vec{p})w^\mp_{\vec{k}_+} w^\pm_{-\vec{p}},\label{eq:fourierapprox}
\end{align}
where \m{\vec{k}_{\pm}\doteq \vec{k} \pm \vec{p}} and \m{\vec{k}= \vec{q}+n\vec{p}} with integer \m{n}. For all other \m{\vec{k}}, one obtains \m{\pd_t w^\pm_{\vec{k}} \approx 0}. In particular, this means that \m{w^\pm_{\vec{p}}} can be considered fixed; \ie \m{A^\pm\doteq w^\pm_{\vec{p}} \approx \const}. In terms of the Els\"asser fields, this corresponds to a primary structure
\begin{gather}
\label{eq:prifields}
\vec{z}^\pm = -\frac{2}{p}\,\mc{A}^\pm \sin(py + \theta^\pm)\uvec{x}, 
\end{gather}
where \m{\mc{A^\pm} \doteq |A^\pm|} and \m{\theta^\pm \doteq \arg A^\pm}.

For the modulation energy density and the primary-wave energy density,
\begin{subequations}\label{eq:energies0}
\begin{gather}
E_{\text{mod}} \doteq \sum_{\vec{k}=\vec{q}+n\vec{p}} E_{\vec{k}}=\sum_n \frac{\rho}{4}\sum_{\sigma=\pm1} 
\frac{|w_{\vec{q}+n\vec{p}}^{\sigma}|^2}{q^2 + n^2p^2},
\label{eq:Emod0}
\\
E_{\text{pri}} \doteq E_{\vec{p}}=\frac{\rho}{4}\sum_{\sigma=\pm1} \frac{|w_{\vec{p}}^{\sigma} |^2}{p^2},
\label{eq:Epri0}
\end{gather}
\end{subequations}
one has
\begin{subequations}\label{eq:energydyn}
\begin{gather}
\frac{\dd E_{\text{mod}}}{\dd t} = \sum_n \frac{\rho}{4}\sum_{\sigma=\pm1} 
\frac{w_{\vec{q} + n\vec{p}}^{\sigma*}\, \pd_t w_{\vec{q} + n\vec{p}}^{\sigma}}{q^2 + n^2p^2}
+ \text{c.c.},\label{eq:energydynmod}
\\
\frac{\dd E_{\text{pri}}}{\dd t} = \frac{\rho}{4}\sum_{\sigma=\pm1} \frac{w_{\vec{p}}^{\sigma*}\dot w_{\vec{p}}^{\sigma}}{p^2}  + \text{c.c.},
\label{eq:energydynpri}
\end{gather}
\end{subequations}
where `c.c.' stands for `complex conjugate'. Using 
\begin{gather}\label{eq:dynpri}
\pd_t w_{\vec{p}}^{\sigma}=\sum_{\vec{k}=\vec{q}+n\vec{p}} T(\vec{k},\vec{p-k}) w_{\vec{k}}^{-\sigma} w_{\vec{p-k}}^\sigma,
\end{gather}
for \m{\vec{k}=\vec{q}+n\vec{p}}, one finds that our approximate equations conserve the total energy density \m{E_{\text{tot}} = E_{\text{pri}} + E_{\text{mod}}} exactly within this approximation. (For brevity, from now on we will refer to the energy densities simply as energies.) We will assume that the initial modulational energy is negligible, so \m{E_{\text{tot}} = \rho\mc{A}^2/4p^2}, where \m{\mc{A}^2\doteq(\mc{A}^+)^2+(\mc{A}^-)^2}, and we will also use the following dimensionless energies:
\begin{subequations}
    \begin{gather}
        \mc{E}_{\text{pri}}=E_{\text{pri}}/E_{\text{tot}},\label{eq:normEpri}\\
        \mc{E}_{\text{mod}}=E_{\text{mod}}/E_{\text{tot}},\label{eq:normEmod}\\
        \mc{E}_v=E_v/E_{\text{tot}},\label{eq:normEv}\\
        \mc{E}_v=E_b/E_{\text{tot}}\label{eq:normEb}.
    \end{gather}
\end{subequations}

\subsubsection{Dimensionless equations}
\label{sec:dimeq}

In order to reduce the number of free parameters, let us perform a variable transformation \m{w_{\vec{q}+n\vec{p}}^\pm \mapsto \amp_n^\pm}:
\begin{gather}
w_{\vec{q}+n\vec{p}}^\pm = 
\sqrt{\frac{\mc{A}^3(q^2+n^2p^2)}{\mc{A}^\mp p^2}}
\exp\left(\ii n\, \frac{\theta^+ + \theta^-}{2}\right) \amp_n^\pm,
\end{gather}
where \m{\amp_n^\pm = \mc{O}(\epsilon)}. Then, \Eq{eq:fourierapprox} becomes
\begin{align}\label{eq:intermediate}
\pd_t \amp_n^\pm = \quad&\mc{A}^\mp \ee^{\mp\ii\theta} \alpha_n^- \amp_{n-1}^\pm 
+ \sqrt{\mc{A}^+\mc{A}^-}\ee^{\pm \ii\theta}\beta_n^- \amp_{n-1}^\mp\notag\\
+ & \mc{A^\mp}\ee^{\pm\ii\theta}\alpha_n^+ \amp_{n+1}^\pm 
+ \sqrt{\mc{A}^+\mc{A}^-}\ee^{\mp \ii\theta}\beta_n^+ \amp_{n+1}^\mp,
\end{align}
where \m{\theta\doteq(\theta^+-\theta^-)/2} and \m{\alpha_n^\pm} and \m{\beta_n^\pm} are given by
\begin{subequations}
\label{eq:ab}
\begin{gather}
 \alpha_n^\pm \doteq \mp r\,\frac{r^2 + n(n \pm 1)}{\sqrt{(r^2+n^2)(r^2 + (n \pm 1)^2})},
 \\
 \beta_n^\pm \doteq -r\,\frac{n}{\sqrt{(r^2+n^2)(r^2 + (n \pm 1)^2)}}.
\end{gather}
\end{subequations}
Note that these coefficients depend only on the ratio \m{r \doteq q/p} rather than on \m{q} and \m{p} separately. From \Eq{eq:intermediate}, one can also see that the individual phases \m{\theta^+} and \m{\theta^-} per se are, expectedly, unimportant  and the dynamics is affected by these phases only in the combination \m{\theta^+ - \theta^-}.

It is also convenient for our purposes to introduce
\begin{gather}
\tau \doteq \mc{A} \,t
\end{gather}
as the new dimensionless time, as well as \m{\phi \doteq \arctan(\mc{A}^-/\mc{A}^+) \in [0, \upi/2]}, so that
\begin{gather}
\mc{A}^+ = \mc{A}\cos\phi, 
\qquad
\mc{A}^- = \mc{A}\sin\phi.
\end{gather}
In particular, \m{\phi = 0} and \m{\phi = \upi/2} correspond to structures consisting only of \m{w_{\vec{p}}^+} and \m{w_{\vec{p}}^-}, respectively, and \m{\phi = \upi/4} corresponds to a `balanced' background with \m{\mc{A}^+ = \mc{A}^-}. For simplicity, \m{\phi = \upi/4} will be assumed from now on throughout the paper (except where specified otherwise). In this case, one has
\begin{gather}\label{eq:thetameaning}
\frac{E_{\text{pri}, b}}{E_{\text{pri}, v}}
= \frac{|w_{\vec{p}}^+ -w_{\vec{p}}^-|^2}{|w_{\vec{p}}^+ +w_{\vec{p}}^-|^2}
= \tan^2\theta,
\end{gather}
where \m{E_{\text{pri}, v}} and \m{E_{\text{pri}, b}} are the kinetic and magnetic energy densities in the primary mode respectively.  Hence, \m{\theta} can be understood as the parameter that determines the relative weights of the kinetic and magnetic-field energy in the primary-mode energy. In particular, \m{\theta = 0} corresponds to a primary structure that involves only velocity perturbations, while \m{\theta = \upi/2} corresponds to a primary structure that involves only magnetic-field perturbations.

Using this notation, one can express \Eq{eq:intermediate} in the following compact form:
\begin{gather}
\label{eq:general}
\dot{\vec{\amp}}_n = \sum_{\sigma = \pm 1}\vec{G}_{n}^\sigma\vec{\amp}_{n+\sigma}
\end{gather}
(the dot denotes \m{\pd_\tau}), or even more compactly as 
\begin{gather}\label{eq:matrixeq}
\dot{\vec{\amp}} = \matr{M}\vec{\amp}.
\end{gather}
Here, \m{\vec{\amp}_n \doteq (\amp_n^-,  \amp_n^+)^\intercal} is a two-component column vector (the symbol \m{^\intercal} denotes transposition), \m{\vec{\amp} \doteq (\hdots, \vec{\amp}_{-1}, \vec{\amp}_0,\vec{\amp}_1 \hdots)^\intercal} is an infinite-dimensional block vector consisting of \m{\vec{\amp}_n}, the matrices \m{\matr{G}_{n}^\sigma} are given~by
\begin{gather}
\matr{G}_{n}^\sigma \doteq 
 \begin{pmatrix}
  \alpha_n^\sigma\cos\phi \,\ee^{-\ii\sigma\theta}
  &
  \beta_n^\sigma\sqrt{\cos\phi\sin\phi}\,\ee^{\ii\sigma\theta}
  \\
  \beta_n^\sigma \sqrt{\cos\phi\sin\phi}\,\ee^{-\ii\sigma\theta}
  &
  \alpha_n^\sigma\sin\phi \,\ee^{\ii\sigma\theta}
 \end{pmatrix},
\end{gather}
and \m{\matr{M}} is the following block matrix:
\begin{equation*} 
 \vec{M}\doteq\begin{mpmatrix}\ddots& & &\vdots& & &\iddots\\
    &0&\vec{G}_{-2}^+&0&0&0& \\
   &\vec{G}_{-1}^-&0&\vec{G}_{-1}^+&0&0& \\
   \hdots&0&\vec{G}_0^-&0&\vec{G}_0^+&0&\hdots\\
   &0&0&\vec{G}_1^-&0&\vec{G}_1^+\\
      &0&0&0&\vec{G}_2^-&0&\\
   \iddots& & &\vdots& & &\ddots
 \end{mpmatrix}.
\end{equation*}
It is readily seen then that the system's dynamics is controlled only by the dimensionless parameters \m{r}, \m{\theta}, and \m{\phi}, while \m{\mc{A}} and \m{p} determine only its characteristic temporal and spatial scales, respectively.

Equations \eq{eq:general} and \eq{eq:matrixeq} are similar to the equations that describe a one-dimensional chain of coupled linear oscillators. The \m{n}th elementary cell of the chain consists of two different oscillators characterised by \m{\amp_n^+} and \m{\amp_n^-}, respectively. Similar equations emerge in studies of phase-space turbulence for the coupling between Hermite moments of the particle phase-space probability distribution \citep{ref:hammett93, ref:nastac23, ref:adkins18, ref:jtparker16, ref:kanekar15}. Throughout this work, we use \Eqs{eq:general} and \eq{eq:matrixeq} to study the nature of collective oscillations in the chain of modulational harmonics, which can be understood as Floquet modes of the linearised system. Unstable oscillations of this system (\ie MIs) lead to structure formation on top of the primary structure.

One can also consider \Eq{eq:matrixeq} as a Schr\"odinger equation with a Hamiltonian \m{\ii\matr{M}}. This Hamiltonian is not Hermitian, because modulations are parametrically coupled with the primary mode (through which energy can be either gained or lost) and dissipate at \m{n\to\infty}. At the same time, this Hamiltonian is invariant under the time-reversal transformation (\m{\ii\to-\ii}) and the parity transformation in the spectral space (\m{n\rightarrow -n}, \m{\sigma\rightarrow-\sigma}). This makes the system \eq{eq:general} \m{\mc{PT}}-symmetric \citep{ref:bender05,book:bender19}. Depending on the balance of sources and sinks, such systems can support modes with entirely real frequencies (unbroken \m{\mc{PT}} symmetry) and pairs of modes whose frequencies are mutually complex-conjugate (broken \m{\mc{PT}} symmetry). This will be discussed further in \Sec{sec:spectralw}.

\begin{figure}
 \centering
 \includegraphics{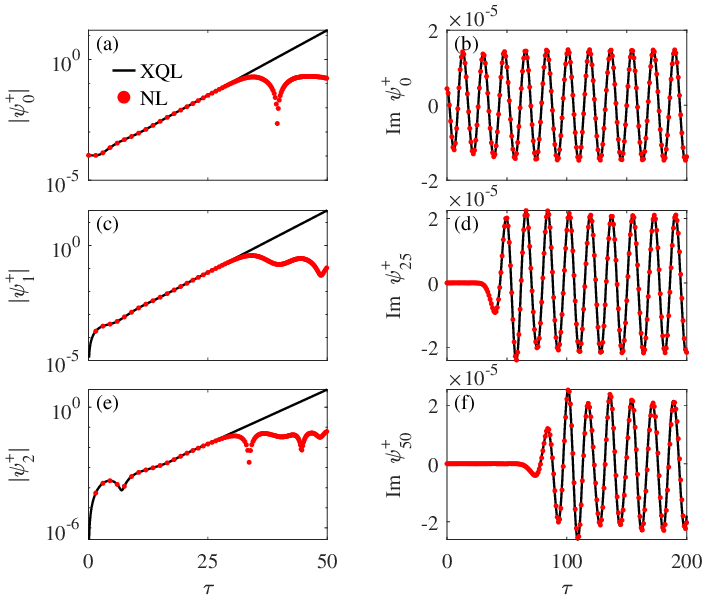}
 \caption{\label{fig:XQLTNLcomp}
Comparison of the time evolution of various \m{\psi_n^+} as predicted by DNS of the nonlinear equation \eq{eq:elsasser} (red dots) and XQLT \eq{eq:general} (black curves) for two representative cases of the MI (\m{\theta = \upi/6}) [(a), (c), (e)] and stable modulational dynamics (\m{\theta = \upi/3}) [(b), (d), (f)]. Nonlinear DNS are seeded with random noise, resulting in a broadband modulational dynamics, but the results shown are for a modulation corresponding to \m{r = 0.5}, and the XQLT simulations are initialised accordingly. For the \m{\theta = \upi/6} case, the harmonics shown are: (a) \m{n = 0}, (b) \m{n = 1}, and (c) \m{n = 2}. In this case, XQLT adequately approximates the nonlinear dynamics until about \m{\tau \sim 30}. For the \m{\theta =  \upi/3} case, the harmonics shown are: (d) \m{n = 0}, (e) \m{n = 25}, and (f) \m{n = 50}. In this case, XQLT adequately approximates the nonlinear dynamics indefinitely.}
\end{figure}

As long as the underlying ordering assumption [\m{|w_{\vec{q}+n\vec{p}}^\pm|/|w_{\vec{p}}^\pm| \sim \Obig(\epsilon)}] holds, XQLT \eq{eq:general} provides an excellent approximation of the nonlinear modulational dynamics of \Eq{eq:elsasser} (\Fig{fig:XQLTNLcomp}). However, practical applications of XQLT require truncations of this system. In the remainder of this work, we discuss to what extent such truncations can be adequate and compare our analytical results to DNS of \Eq{eq:general}, in lieu of the full \Eq{eq:elsasser}.

\section{Truncated models}
\label{sec:trunc}

\subsection{Basic equations}
\label{sec:truncb}

A common way to truncate a problem like \Eq{eq:matrixeq} is to postulate that all \m{\amp_n} with \m{|n| > N} are negligible. One can also understand this as imposing reflective boundary conditions in the oscillator chain \m{\{\vec{\amp}_n^\sigma\}}:
\begin{gather}\label{eq:bc}
\vec{\amp}_{N+1}^\sigma = \vec{\amp}_{-(N+1)}^\sigma = 0.
\end{gather}
Retained in this case are harmonics with wavevectors \m{\vec{p}} and \m{\vec{q} + n\vec{p}}, with \(n = 0, \pm 1, \ldots , \pm N\). This means that the resulting system has \m{Q = 1 + (1 + 2N)} degrees of freedom. We call this procedure \m{Q}-mode truncation (\m{Q}MT). The corresponding truncation of \m{\vec{\amp}} will be denoted as \m{\vec{\amp}_{Q\text{MT}}}, and the corresponding truncation of \m{\matr{M}} will be denoted as \m{\matr{M}_{Q\text{MT}}}, so \Eq{eq:matrixeq} becomes
\begin{gather}\label{eq:truncated1}
\dot{\vec{\amp}}_{Q\text{MT}} = \matr{M}_{Q\text{MT}}\vec{\amp}_{Q\text{MT}}.
\end{gather}
Because of \Eq{eq:bc}, eigenmodes of this system can be understood as standing waves in the oscillator chain \m{\{\vec{\amp}_n^\sigma\}} and have the form \m{\vec{\amp}_{Q\text{MT}} = \vec{Y}\exp(-\ii\omega \tau)}, where \m{\omega} are global frequencies and \m{\vec{Y}} are constant `polarization vectors'. According to \Eq{eq:matrixeq}, these vectors satisfy \m{\matr{D}\vec{Y} = 0}, where \m{\matr{D} = \ii \omega \mathbb{1} + \matr{M}_{Q\text{MT}}} and \m{\mathbb{1}} is a unit \m{Q\times Q} matrix. Then, \m{\omega} can be found from
\begin{gather}\label{eq:modmodes}
 \det \matr{D} = 0
\end{gather}
and MIs correspond to modes with \m{\im \omega > 0}. 

The success of this approach hinges on the assumption that higher harmonics truly remain negligible. In other words, a \m{Q}MT should be able to adequately describe the evolution of a modulational mode if its eigenvector \m{\vec{\amp}_{Q\text{MT}}} is heavily weighted towards the low-order harmonics. Below, we explore to what extent this assumption is satisfied in various regimes.

\subsection{Growth rates}

The lowest-order \m{Q}MT is 4MT, which corresponds to \m{N = 1}. Within this model, the primary harmonic with \m{\vec{k} = \vec{p}} is assumed to interact with the modulational mode at \m{\vec{k} = \vec{q}} only through two sidebands \m{\vec{k} = \vec{q} \pm \vec{p}}; then, \m{\vec{\amp}_\text{4MT} = (\vec{\amp}_{-1}, \vec{\amp}_0, \vec{\amp}_1)^\intercal} and
\begin{gather}\label{eq:4mt}
\matr{M}_{\text{4MT}} =
\begin{mpmatrix}
 0&\vec{G}_{-1}^+&0 \\
 \vec{G}_0^-&0&\vec{G}_0^+\\
 0&\vec{G}_1^-&0
\end{mpmatrix}.
\end{gather}
(This model can be understood as quasilinear, because, although the modulational dynamics remains nonlinear, the second- and higher-order harmonics of the perturbation are neglected.) The 4MT equations yield
\begin{gather}\label{eq:4mtw}
 \omega^2 = \frac{r^4}{1+r^2}\left(1 \pm \sqrt{1 - (1 - r^{-4} \cos^22\theta) \sin^2 2\phi}\right).
\end{gather}
[Note that these are normalised frequencies corresponding to the normalised time \m{\tau}. To obtain the actual physical frequencies, one must multiply the right-hand side of \Eq{eq:4mtw} by \m{\mc{A}}.] Assuming \m{\phi = \upi/4}, one finds that two out of four branches of this dispersion relation are unstable at \m{|r| < 1}. The corresponding growth rate \m{\Gamma \doteq \im\omega} is maximised at \m{\theta = 0} and \m{\theta = \upi/2} at the value
\begin{gather}\label{eq:4mtg}
 \Gamma = r \sqrt{\frac{1 - r^2}{1 + r^2}}
\end{gather}
(from now on, we assume \m{r > 0} for clarity of notation), and the corresponding polarization vectors are as follows:
\begin{gather}\label{eq:polarization}
\vec{Y}_0 = 
 \begin{cases}
  (1, +1)^\intercal, & \theta = 0,\\
  (1, -1)^\intercal, & \theta = \upi/2.\\        
 \end{cases}
\end{gather}
The case \m{\theta = 0} can be recognised as a purely hydrodynamic Kelvin--Helmholtz instability (KHI), \ie one that does not involve magnetic field. The case \m{0 < \theta < \upi/4} can be understood as the MHD generalization of the KHI. 

Potentially more accurate is a model with \m{N = 2}, or 6MT, which accounts for harmonics with \m{\vec{k} = \vec{q} \pm 2\vec{p}}; then, \m{\vec{\amp}_\text{4MT} = (\vec{\amp}_{-2}, \vec{\amp}_{-1}, \vec{\amp}_0, \vec{\amp}_1, \vec{\amp}_{2})^\intercal} and
\begin{gather}\label{eq:6mt}
\matr{M}_{\text{6MT}} =
\begin{mpmatrix}
 0&\vec{G}_{-2}^+&0&0&0\\
 \vec{G}_{-1}^-&0&\vec{G}_{-1}^+&0 &0\\
 0&\vec{G}_0^-&0&\vec{G}_0^+&0\\
 0&0&\vec{G}_1^-&0&\vec{G}_1^+\\
 0&0&0&\vec{G}_2^-&0
\end{mpmatrix}.
\end{gather}
One can derive an analytical expression for \m{\omega} from here just like we derived \Eq{eq:4mtw} from \Eq{eq:4mt}. However, this expression is cumbersome and not particularly instructive, so it is not presented explicitly. Truncations with larger \m{Q}, albeit also explored by us, will not be presented either for the same reason.

\begin{figure}
 \centering
 \includegraphics{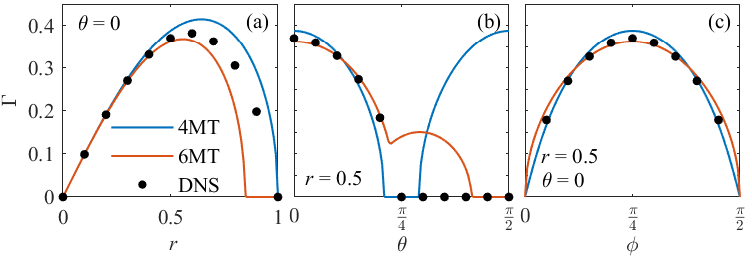}
 \caption{\label{fig:grs}
 (a)~The growth rate \m{\Gamma} (of the most unstable mode) versus the normalised wavenumber \m{r} at \m{\theta = 0}: 4MT (blue), 6MT (red), and nonlinear DNS of \Eq{eq:vecw} (black markers). (b)~Same for \m{\Gamma} versus \m{\phi} at \m{r = 0.5}. The 4MT predicts identical growth rates at \m{\phi = 0} and \m{\phi = \upi}, while the 6MT and nonlinear DNS predict that the system is unstable at \m{\phi = 0} and stable at \m{\phi = \upi}. The growth rates are calculated with nonlinear DNS by adding random noise with \m{k_y = 0} noise to the primary mode, allowing this modulational noise to evolve, and then picking out the growth rate of the most unstable mode. The noise amplitude is taken small enough such that it does not affect the results. }
\end{figure}

Figure \ref{fig:grs} shows a comparison of the 4MT and 6MT predictions and the MI growth rates inferred numerically from nonlinear DNS of \Eq{eq:vecw}. Note that the truncated models quantitatively agree with each other and with the DNS at some parameters but drastically disagree at other parameters. In particular, the 4MT predicts that \m{\Gamma} is an even function of \m{\upi/2 - \theta}, while the 6MT predicts that this is not the case. Furthermore, the DNS predicts that, for example, at \m{r = 0.5}, the system is stable at all \m{\theta > \upi/2} [\Fig{fig:grs}(b)]. Let us discuss this in detail. 

\begin{figure}
 \centering
 \includegraphics{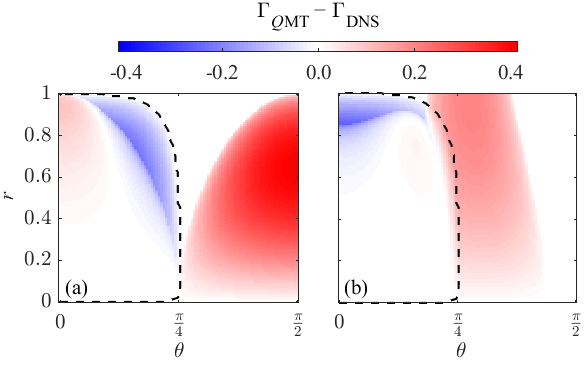}
 \caption{\label{fig:stabmap}
  The difference between the growth rate predicted by a \m{Q}MT and that inferred from XQLT DNS, \m{\Gamma_{Q\text{MT}}-\Gamma_{\text{DNS}}}, versus \m{(\theta, r)} for \m{\phi = \upi/4}: (a)~4MT, (b)~6MT. The dashed curves mark the stability boundary as determined by a parameter scan with nonlinear DNS. The preponderance of the red colour, especially outside the instability domain, indicates that truncated models tend to produce false positives for instability.
 }
\end{figure}

\begin{figure}
 \centering
 \includegraphics{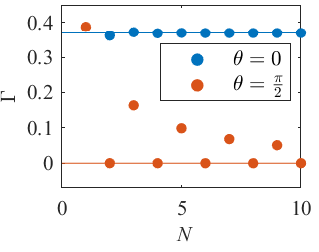}
 \caption{\label{fig:truncationeffects}
  The growth rates \m{\Gamma} obtained through a \m{Q}MT for \m{\theta = 0} (blue markers) and \m{\theta = \upi/2} (orange markers) versus the number of modes retained, \m{N}. The solid coloured lines indicate the corresponding DNS growth rates, to which the rates predicted by the truncated models converge at large \m{N}. 
 }
\end{figure}

As expected from the analytical formula \eq{eq:4mtg} based on the 4MT, and as we have also found numerically, MI is typically suppressed at \m{r \gtrsim 1}; hence, below we focus on the regime \m{r \in (0,1)}. A comparison of the growth rate predicted by nonlinear DNS and truncated models in this range is presented \Fig{fig:truncationeffects}. The corresponding primary modes with \m{\theta \lesssim \upi/4} are modulationally unstable, and the growth rates predicted by 4MT and 6MT are in good agreement with DNS. Such primary structures are dominated by the velocity field, \m{\mc{E}_{\text{pri}, v} > \mc{E}_{\text{pri}, b}} [see \eq{eq:thetameaning}], and thus will be called \m{\vec{v}}-dominated primary modes (VDPMs). At \m{\theta \gtrsim \upi/4}, primary structures are dominated by the magnetic field, \m{\mc{E}_{\text{pri}, b} > \mc{E}_{\text{pri}, v}}, and thus will be called \m{\vec{b}}-dominated primary modes (BDPMs). As seen in \Fig{fig:truncationeffects}, BDPMs are modulationally stable, but truncated models predict otherwise. Figures \ref{fig:stabmap}(a) and (b) show the discrepancy between the predictions of nonlinear DNS, 4MT, and 6MT for representative parameters, specifically, \m{\theta \in (0, \upi/2)} and \m{\phi = \upi/4}. (Stability is primarily determined by \m{\theta} and \m{r}. Deviations of \m{\phi} from \m{\upi/4} result only in quantitative adjustments, unless \m{\phi} is close to~\m{0} or~\m{\upi/2}.) In other words, truncated modes tend to overestimate the growth rate and systematically produce false positives for instability. Figure~\ref{fig:truncationeffects} shows that, although the magnitude of disagreement decreases with \m{N}, the systematic overprediction of instability persists.

The reason for this overprediction can be understood by analyzing the eigenmode structure. (See also \App{app:timescales} for an alternative explanation.) As to be discussed in \Sec{sec:spectralw}, unstable modes are \textit{evanescent spectral waves}, whose \m{|Y_n|} exponentially decrease with \m{|n|} [\Fig{fig:eigens}(a)]:
\begin{gather}\label{eq:Ynu}
|Y_n| \propto \exp(-|n|/l).
\end{gather}
(We assume \m{\phi = \upi/4} for simplicity, so \m{|Y_n^+| = |Y_n^-|} and the upper index in \m{|Y_n^\sigma|} can be omitted.) That is why truncated models correctly predict MI for modes that are actually unstable. The spectral scale~\m{l} depends on the system parameters, and the smaller \m{l} the more accurate \m{Q}MT is. At some parameters, though, particularly when \m{\theta \rightarrow \upi/4}, \m{l} becomes large or even infinite, such that \m{|Y_n|} asymptotes to a nonzero constant at \m{|n| \to \infty} [\Fig{fig:eigens}(b)]. In this case, \m{Q}MT is bound to fail. The corresponding modes are \textit{propagating spectral waves} (PSWs). While they also receive energy from the primary structure at low \m{n}, PSWs transport this energy along the spectrum to \m{|n| \to \infty}. There, the energy is unavoidably dissipated by viscosity or resistivity, however small those are. This dissipation makes PSWs more stable than any \m{Q}MT would predict, because \m{Q}MTs assume that the mode energy forever resides at small \m{n}, where dissipation is small or zero. Below, we discuss these effects in more detail.

\section{Spectral waves}
\label{sec:spectralw}

\subsection{Dynamics at large \texorpdfstring{$|n|$}{|n|}}

First, let us consider spectral waves at \m{|n| \gg 1}. In this limit, one has \m{\alpha_n^\pm\rightarrow\mp r} and \m{\beta_n^\pm\rightarrow 0}, so \Eq{eq:matrixeq} yields two decoupled wave equations for~\m{\amp_n^+} and~\m{\amp_n^-}:
\begin{subequations}\label{eq:waveeqs}
\begin{align}
 \dot{\amp}_n^-
  & = r\cos\phi \big(\ee^{\ii\theta} \amp_{n-1}^- - \ee^{-\ii\theta}\amp_{n+1}^-\big),\\
 \dot{\amp}_n^+
 & = r\sin\phi \big(\ee^{-\ii\theta} \amp_{n-1}^+ - \ee^{\ii\theta}\amp_{n+1}^+\big).
\end{align}
\end{subequations}
These equations allow solutions in the form of monochromatic waves:
\begin{gather}
\amp_n^\sigma = \Psi^\sigma \exp(- \ii\omega \tau + \ii K^\sigma n),
\end{gather}
with constant amplitude \m{\Psi^\sigma}, frequency \m{\omega}, and `wavenumber' \m{K^\sigma}. The quotation marks are added as a reminder that the corresponding waves propagate in the spectral space rather than in the physical space. Also note that \m{K^\pm} are defined unambiguously only within the first Brillouin zone, \m{K^\pm \in (-\upi, \upi)}.

\begin{figure}
 \centering
 \includegraphics[width=.9\linewidth]{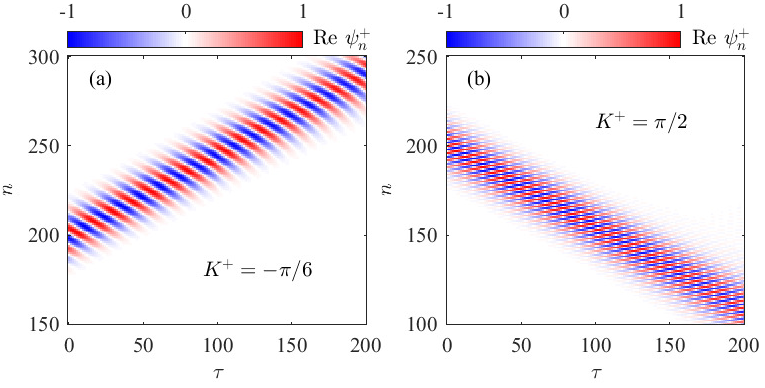}
 \caption{\label{fig:packets}
 Numerical simulations showing PSW packets propagating: (a) down the spectrum (\m{|n|} of the packet center increases with time) and (b) up the spectrum (\m{|n|} of the packet center decreases with time). In both cases, \m{r = 0.5} and \m{\theta = \upi/3}. Both packets are initialised using \m{\psi_n^+(\tau = 0) = \exp[-(n - n_0)^2/\varsigma + \ii K^+ n]} with \m{n_0 = 200}, \m{\varsigma = 150}, and: (a) \m{K^+ = -\upi/6}, (b) \m{K^+ = \upi/2}.
 }
\end{figure}

From \Eqs{eq:waveeqs}, one readily finds that spectral waves obey the following dispersion relations:
\begin{gather}\label{eq:disprel}
 \omega = \omega_0^\sigma \sin \kappa^\sigma,
 \qquad
 \kappa^\pm \doteq K^\pm \pm\theta,
\end{gather}
where \m{\omega_0^+ = 2r\sin\phi} and \m{\omega_0^- = 2r\cos\phi}. Accordingly,spectral waves in the \m{n} space along the \m{\psi^\sigma} channel have the group velocity
\begin{gather}
v_g^\sigma = \omega_0^\sigma \cos \kappa^\sigma
\end{gather}
and can propagate either up or down the spectrum (\Fig{fig:packets}). The dependence of \m{\omega} and \m{v_g^\pm} on \m{\theta} can be removed by a gauge transformation. Specifically, \m{\kappa^\pm} becomes the true wavenumber if one adopts the variables \m{\bar{\amp}_n^\pm \doteq \amp_n^\pm\ee^{\mp\ii n\theta}} instead of \m{\amp_n^\pm}. Also notably, to the extent that the (large) index \m{n} can be approximately considered a continuous variable, \Eqs{eq:waveeqs} can be written as usual nondispersive-wave equations for \m{\zeta^\pm(\tau, n) \doteq \bar{\amp}_n^\pm(\tau)}:
\begin{gather}
 \pd_\tau \zeta^\pm = -\omega_0^\pm\, \pd_n \zeta^\pm.
\end{gather}
This model corresponds to the small-\m{\kappa^\sigma} limit of \Eq{eq:disprel}, that is, \m{\omega \approx \omega_0^\sigma \kappa^\sigma}.

\begin{figure}
 \centering
 \includegraphics{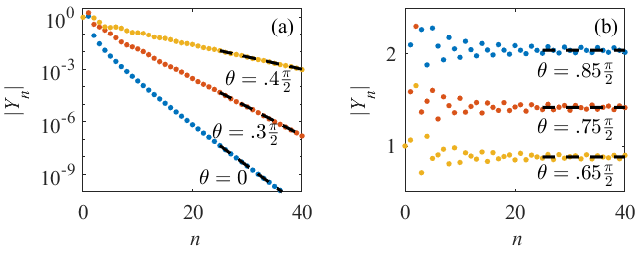}
 \caption{\label{fig:eigens}
 Typical structures of eigenmodes, specifically, \m{|Y_n|} versus \m{n}, at \m{r=0.5} and \m{\phi=\upi/4} for various \m{\theta}: (a)~unstable modes supported by VDPMs; (b)~oscillatory solutions supported by BDPMs. (The upper index in \m{|Y_n^\sigma|} is omitted because \m{|Y_n^+| = |Y_n^-|} for the assumed parameters.) \m{|Y_n|} decreases exponentially with \m{n} for the former (notice the logarithmic scale) but asymptote to nonzero constants for the latter. The asymptotes are indicated by the dotted lines. These results are obtained through DNS of \Eq{eq:matrixeq} using the initial conditions of the form \Eq{eq:ic} and normalised such that \m{|Y_0| = 1}. }
\end{figure}

The above equations can be used to describe PSW packets localised at large \m{|n|} but not the global eigenmodes (because the latter involve dynamics at small \m{|n|}). In particular, \Eqs{eq:disprel} are not enough to find the global-mode frequencies. Yet they allow one to determine the mode structure at \m{|n| \gg 1} if one knows the mode frequency from other considerations (or, vice versa, to find the mode frequency if \m{\kappa^\pm} are known). In particular, unstable modes have complex \m{\omega}, so, according to \Eq{eq:disprel}, their asymptotic wavenumber cannot be real. This explains why unstable modes are evanescent. In contrast, for real~\m{\omega}, \Eq{eq:disprel} allows for real wavenumbers (provided that \m{|\omega| \le \omega_0^\pm}), \ie predicts PSWs. These results are corroborated by DNS of \Eq{eq:matrixeq}. For example, \Fig{fig:eigens} shows the mode structures inferred from the asymptotic dynamics of \m{\amp_{n}^\sigma(\tau)} at large enough \m{\tau} for the initial conditions
\begin{gather}\label{eq:ic}
 \amp_{n}^\sigma(\tau=0) =
 \begin{cases}
  \epsilon \exp(\ii\xi^\sigma_n), & n = \pm 1,\\
  0, & n \neq \pm 1. 
 \end{cases}
\end{gather}
Here, \m{\epsilon} is a constant small amplitude (within the linear approximation, one can always rescale \m{\amp_{n}^\sigma} such that \m{\epsilon = 1}), and \m{\xi_n} are parameters that change the polarization of the initial conditions while keeping the initial modulation energy fixed. (The specific values of \m{\xi_n^\sigma} are given in the captions of figures in which initial condition dependent results of DNS are presented.) This form of the initial conditions is chosen to emulate a simple modulation of the primary mode.

The fact that the properties of spectral waves are determined only by the asymptotic properties of \m{\alpha_n^\pm} and \m{\beta_n^\pm} makes these waves particularly robust. In particular, note that the asymptotic form of \m{\alpha_n^\pm} and \m{\beta_n^\pm} at \m{|n| \to \infty} remains the same even when \m{\vec{p}} and \m{\vec{q}} are not orthogonal, so the above results readily extend to general \m{\vec{q}}. Likewise, a similar picture holds also for three-dimensional interactions, although the mode polarization can be more complicated in this case. Notably, PSWs provide ballistic rather than diffusive (or super-diffusive) energy transport along the spectrum. This distinguishes them from the cascades that are commonly discussed in turbulence theories and result from eddy--eddy interactions, or wave--wave collisions \citep{ref:galtier00}.

\subsection{Global modes in the form of PSWs}
\label{sec:global}

In realistic settings, spectral waves are excited at finite \m{|n|} and propagate down the spectrum. Evanescent waves reach \m{|n| \sim l} [see \Eq{eq:Ynu}] in finite time \m{\tau_0}, get reflected, and eventually (asymptotically) settle as stationary standing waves on time scales of the order of a few \m{\tau_0}. Such waves can be adequately described by truncated models. In contrast, PSWs continue to propagate toward infinity indefinitely (until they dissipate at scales where viscosity or resistivity is no longer negligible). Thus, they never develop components propagating up the spectrum at large \m{|n|} and so they \textit{never} become quite like the standing-wave eigenmodes predicted by truncated models (\Sec{sec:truncb}). This means that actual PSWs cannot be adequately described by analyzing \m{Q}MT in principle. Instead, they should be considered in the context of the initial-value problem, much like Langmuir waves appear in plasma kinetic theory within Landau's initial-value approach as opposed to the Case--van~Kampen true-eigenmode approach \citep{book:stix}. 

\begin{figure}
 \centering
\includegraphics{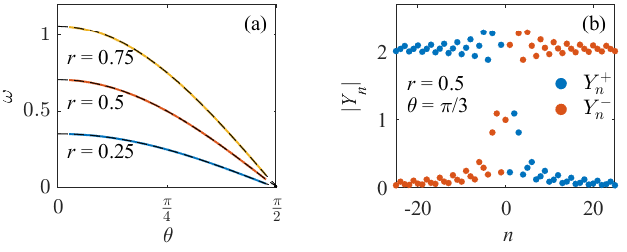}
 \caption{\label{fig:psweigenmodes}
(a)~The dependence of the global-mode frequency \m{\omega > 0} on \m{\theta} for various \m{r}. The dashed lines are the inferred solutions \eq{eq:inferredsol}. (b)~\m{|Y_n^+|} (blue) and \m{|Y_n^-|} (orange) for the mode with \m{\omega > 0}. The eigenmode amplitudes for the \m{\omega < 0} mode are identical with the roles of \m{|Y_n^+|} and \m{|Y_n^-|} switched.
 }
\end{figure}

\begin{figure}
 \centering
 \includegraphics{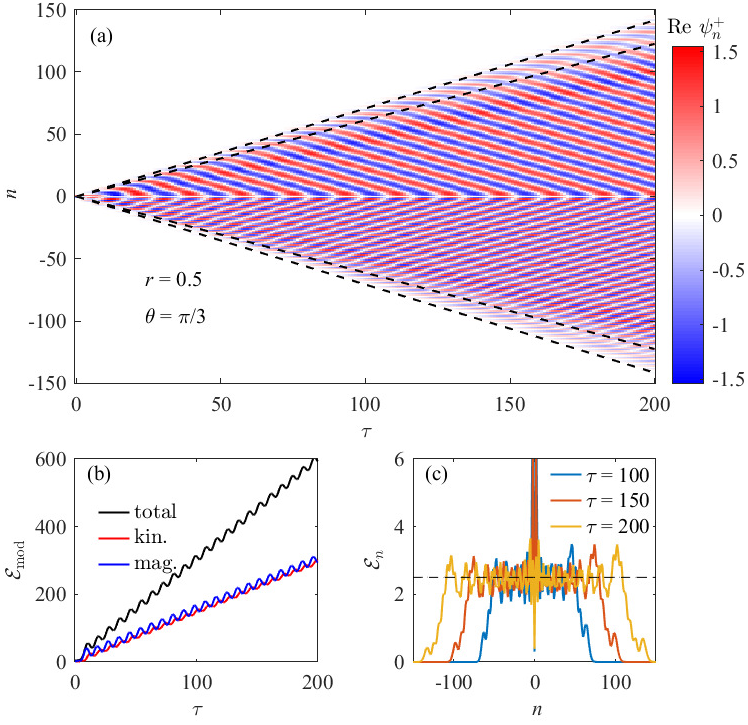}
 \caption{\label{fig:psws}
 DNS of \Eq{eq:matrixeq} showing global-mode PSWs for \m{r=0.5}, \m{\theta = \upi/3}, and \m{\phi = \upi/4}: (a)~\m{\re \amp_n^+/\epsilon} for a PSW seeded by the initial conditions \eq{eq:ic} with \m{\xi^\sigma_d = d\exp(\sigma \ii \upi/4)} (colour bar).
 The dashed lines indicate the fronts propagating at: (i)~the maximum spectral speed \m{\sqrt{2}r} and (ii)~the actual group velocity of the mode. The field between these dashed lines consists of transients, which are negligible behind the second front. (b)~The total normalised energy of the modulation, \m{\mc{E}_{\text{mod}}/\epsilon^2}, versus \m{\tau} (black), along with its kinetic (red) and magnetic (blue) components. (c)~The profiles of the spectral energy density at \m{\tau=100,\,150,\,200}, clearly exhibiting left- and right-propagating fronts. The horizontal dashed line indicates the average spectral energy density \m{\overbar{\mc{E}_n}}, where the average is taken over the PSW period and over \m{n} between the energy fronts.
 }
\end{figure}

Consequently, even though \Eq{eq:matrixeq} has infinitely many degrees of freedom, the number of \textit{relevant} global modes, \ie those that propagate towards infinity, is finite. We find that there are only two of them, which can be attributed to the fact that \m{\dim\vec{\amp}_n = 2}. (Likewise, at a given wavenumber, only one, Langmuir, branch of relevant electrostatic waves in electron plasma remains relevant after transients are gone.) For example, for \m{\phi = \upi/4}, we find from DNS that
\begin{gather}\label{eq:inferredsol}
 \omega = \pm \sqrt{2}r \cos \theta.
\end{gather}
This dependence and the corresponding mode structures are illustrated in \Fig{fig:psweigenmodes}. Each of the two modes has two wavenumbers associated with it: one determines \m{\psi^\sigma_n} at \m{n \to +\infty}, and the other one determines \m{\psi^{-\sigma}_n} at \m{n \to -\infty} (\Fig{fig:psws}). Thus, there are four \m{\kappa^\sigma_d} overall, where \m{\sigma} denotes the corresponding component of \m{\vec{\psi}_n}, and \m{d} denotes the direction of propagation (\m{d = \pm 1} is the sign of the group velocity at \m{|n| \to \infty}). They can be found by combining \Eq{eq:inferredsol} with \Eq{eq:disprel} and applying \m{\sgn v_g = d}, \ie \m{\omega_0^\sigma d\cos \kappa^\sigma_d > 0}. This yields
\begin{subequations}\label{eq:omegas}
 \begin{alignat}{3}
  & \omega > 0: 
  \quad
  && \kappa^+_+=\frac{\upi}{2}-\theta,
  \kern 10pt
  & & \kappa^-_-=-\frac{3\upi}{2}+\theta,
  \\
  & \omega < 0:
  \qquad 
  && \kappa^+_-=\frac{3\upi}{2}-\theta,
  \kern 10pt
  & & \kappa^-_+=-\frac{\upi}{2}+\theta,
 \end{alignat}
\end{subequations}
where \m{\kappa^\sigma_d=K^\sigma_d+\sigma \theta} and the values of \m{K^\sigma_d} are defined modulo the Brillouin zone (\ie \m{K^-_-=-3\upi/2} should be understood as \m{\upi/2} and so on). Remember, though, that this applies only to global eigenmodes. Transient waves propagating in the region \m{|n| \gg 1} can have any \m{\kappa^\sigma} (\Fig{fig:packets}).


\begin{figure}
 \centering
 \includegraphics{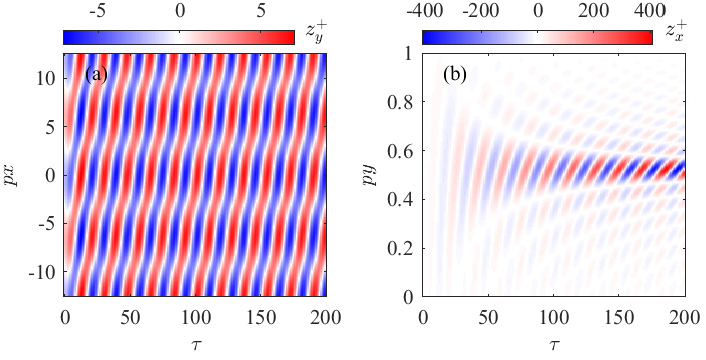}
 \caption{\label{fig:pswinspace}
  The global-mode PSW mode with \m{\omega > 0}, \m{r = 0.5}, and \m{\theta = \upi/3}: same as \Fig{fig:psws}, but in the real space as opposed to the spectral space. Specifically shown are: (a) \m{z_y^+} as a function of \m{(\tau, x)} at \m{y = 0}, (b) \m{z_x^+} as a function of \m{(\tau, y)} at \m{x = 0}.
 }
\end{figure}

The fact that the frequencies of these modes are real indicates that the modes are self-sustained notwithstanding the dissipation that they unavoidably experience at \m{|n| \to \infty}. The corresponding evolution of the wave spectrum is illustrated in \Fig{fig:psws}(a) and the corresponding evolution in the physical space is shown in \Fig{fig:pswinspace}. This remarkable dynamics is understood from the fact that the energy drain via this dissipation is balanced by the energy injection from the primary mode at small \m{|n|}. As seen from \Eqs{eq:energydyn} and \eq{eq:fourierapprox}, the dimensionless modulational energy \eq{eq:normEmod}, or 
\begin{gather}
 \mc{E}_{\text{mod}} \doteq\sum_\sigma \frac{\mc{A}}{\mc{A}^{-\sigma}}\,
 |\amp_n^\sigma|^2
  \equiv \sum_n \mc{E}_n,
\end{gather}
is governed by
\begin{gather}
\dot{\mc{E}}_n=\sum_\sigma (I_n^\sigma+F_n^\sigma) +\text{c.c.},
\end{gather}    
where \m{I_n} and \m{F_n} are given by
\begin{gather}
I_n^\sigma = \frac{\mc{A}\sqrt{\mc{A}^\sigma}}{(\mc{A}^{-\sigma})^{3/2}}\,
(
\beta_n^- \ee^{\ii\sigma\theta} \amp_{n-1}^{-\sigma}
+ \beta_n^+\ee^{-\ii\sigma\theta }\amp_{n+1}^{-\sigma}
)
\amp_n^{\sigma *},
\label{eq:injection}\\
F_n^\sigma = \frac{\mc{A}}{\mc{A}^{-\sigma}}\,
(
\alpha_n^- \ee^{-\ii\sigma\theta} \amp_{n-1}^\sigma 
+ \alpha_n^+ \ee^{\ii\sigma\theta} \amp_{n+1}^\sigma
)
\amp_n^{\sigma*}.
\label{eq:flux}
\end{gather}
Because \m{\sum_n F_n^\sigma = 0}, the terms \m{F_n^\sigma} represent the energy flux that is carried along the modulation spectrum and is conserved within each sub-channel \m{\sigma}. In contrast, \m{I_n} can be understood as injection terms in that they also appear, with the opposite signs, in the equation for the primary wave,
\begin{gather}
 \dot{\mc{E}}_{\text{pri}}
 = -\sum_n \sum_{\sigma=\pm1}I_n^\sigma+\text{c.c.},
\end{gather}
which follows from \Eq{eq:dynpri}. 

The linear spectral waves discussed so far correspond to the regime when the change of \m{\mc{E}_{\text{pri}}} is negligible. Eventually, though, as higher harmonics are populated and thus absorb more energy, the primary wave will be depleted. This means, in particular, that no primary wave can be truly stationary. Even an arbitrarily small perturbation will generally launch a PSW, and this PSW will eventually deplete the primary wave. 

\subsection{Anomalous dissipation}
\label{sec:anomalousdissipation}

Since a PSW mode carries energy towards \m{|n| \rightarrow \infty}, eventually, a large enough \m{|n|} is reached where the energy is dissipated regardless of how small the viscosity and resistivity are. Thus, a PSW exhibits an effective, or `anomalous', dissipation rate \m{\gamma} that is independent of \m{\nu} and \m{\eta} in the limit \m{\nu, \eta \to 0}. This effect is different from the anomalous transport caused by eddy--eddy collisions in turbulence (for example, see \citet{ref:donzis05}) in that the energy transport caused by a PSW is ballistic. As a result, \m{\gamma} is straightforward to calculate, which is done as follows.

\begin{figure}
 \centering
\includegraphics{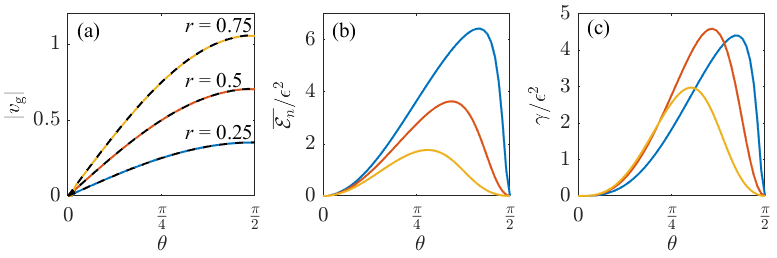}
 \caption{\label{fig:anomdiss}
(a)~The magnitude of the group velocity \m{|v_g|} of the global PSW mode (\ie the speed at which the energy front propagates along the spectrum) versus \m{\theta} for various~\m{r}. (b)~Same for the average spectral energy density \m{\overbar{\mc{E}_n}/\epsilon^2}. (c)~Same for the resulting average drain rate \m{\gamma/\epsilon^2}, where \m{\gamma} is defined in \Eq{eq:dissipationrate}. The initial conditions used throughout these figures are given by \Eq{eq:ic}, with \m{\xi_{1}^\sigma=-\xi_{-1}^\sigma=\exp(\sigma\ii \upi/2).} }
\end{figure}

As discussed in \Sec{sec:global}, a global PSW at large \m{|n|} has a flat mode structure, \ie a structure with \m{\mc{E}_n} independent of \m{n}. This structure establishes itself as an expanding `shelf' whose edges (wave fronts) propagate across the modulational spectrum to \m{n \to \pm \infty} at the group velocity~\m{v_g} (\Fig{fig:anomdiss}). Since the height of the shelf, \m{\overbar{\mc{E}_n}} (the overbar here denotes averaging over the PSW temporal period and over all \m{n} within the shelf), remains constant, this process drains energy from the primary mode linearly in time:
\begin{gather}
\dot{\mc{E}}_{\text{pri}} = - 2|v_g|\overbar{\mc{E}_n} = \const,
\end{gather}
where we consider the dynamics average over the PSW temporal period. Hence, 
\begin{gather}\label{eq:dissipationrate}
\gamma \doteq -\frac{\dot{\mc{E}}_{\text{pri}}}{\mc{E}_{\text{pri}}} 
= 2|v_g|\,\frac{\overbar{\mc{E}_n}}{\mc{E}_{\text{pri}}} \sim \omega_0^\pm \epsilon^2.
\end{gather}
(The rate relative to the physical time \m{t}, as opposed to~\m{\tau}, is \m{\mc{A}} times this \m{\gamma}.)

Let us again assume the initial conditions \eq{eq:ic} and \m{\phi = \upi/4}. Through DNS of \Eq{eq:matrixeq}, we find that the global PSW-mode amplitude is maximised when \m{|\arg(\xi^\sigma_{\pm1}/\xi^{-\sigma}_{\pm1})| = |\arg(\xi^\sigma_{\pm1} / \xi^\sigma_{\mp1})| = \upi}, irrespective of \m{\theta}. (Any particular choice of \m{\xi^\sigma_n} that satisfies this condition affects only phases of the modulational dynamics and does not impact averaged quantities that determine \m{\gamma}.) This corresponds to the case when all modulational-seed energy is in the magnetic field; \ie \m{\mc{E}_b = \mc{E}} and \m{\mc{E}_v = 0}. Figure~\ref{fig:anomdiss} shows the corresponding anomalous dissipation rate normalised to the seed energy, along with its determining factors -- the system-dependent PSW group velocity \m{v_g} and the average spectral energy density \m{\overbar{\mc{E}_n}}, which is determined by the initial conditions. It can be seen that, for the MI unstable VDPMs (\m{\theta < \upi/4}), spectral waves are relatively slow and have a small amplitude. The transition to modulational stability occurs at \m{|v_g| \sim \Gamma}. This condition can be understood as a threshold beyond which (\ie at \m{|v_g| \gtrsim \Gamma}) PSWs provide a sufficiently fast escape route for the energy injected at small \m{|n|}, such that the positive feedback loops [see \eq{eq:injection}] supporting MIs can no longer be sustained. In the context of \m{\mc{PT}} symmetry (\Sec{sec:dimeq}), the spectral group velocity can be understood as the effective coupling between the links in the oscillator chain, connecting the energy injection from the primary mode at small \m{n} to the energy sink at \m{n\to\infty}. As \m{|v_g|} increases with \m{\theta}, the system transitions from broken to unbroken \m{\mc{PT}} symmetry.

Also, notice the following. If modulations at multiple \m{\vec{q}}s are present, they launch independent cascades along \m{\vec{k} = \vec{q} + n\vec{p}} and thus the drain on the shared primary mode will be additive. This regime is illustrated by \Figs{fig:spaceplot}(d)-(f), which show the modulational dynamics for a BDPM (\m{\theta = \upi/3}) seeded with noise. The rather nondescript modulational dynamics seen in \Figs{fig:spaceplot}(d) and~(e) can be understood as the superposition of the many structures like that in \Fig{fig:pswinspace} for various \m{\vec{q}}. Also, the linear-in-time depletion of the primary mode seen in \Fig{fig:spaceplot}(f) can be understood as the sum of the PSW-driven energy fluxes along the constituent modulational channels.

\section{Unified closure}
\label{sec:closure}

\begin{figure}
 \centering
 \includegraphics{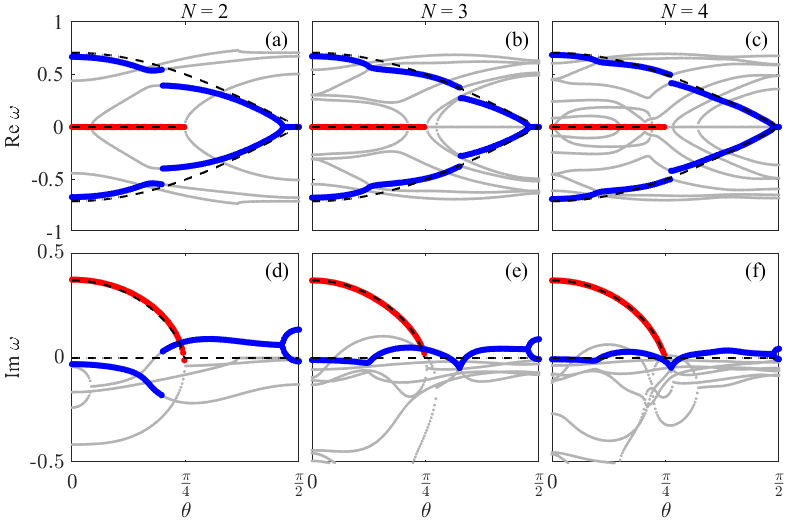}
 \caption{\label{fig:swclosure}
The \m{\theta}-dependence of the global-mode frequencies derived from truncated models with the closure \eq{eq:swclosure} for various \m{N}: \m{\re\omega} (left column) and \m{\im\omega} (right column).~(a) and~(b): \m{N = 2}. (c)~and (d):~\m{N = 3}. (e)~and (f):~\m{N = 4}. The red and blue curves indicate the branches that are the best match to the, respectively, unstable and stable solutions obtained through DNS (black dashed lines). The gray curves show the remaining spurious solutions due to finite \m{N}. Note that MIs exist only for \m{\theta \in (0, \upi/4)}, while PSWs exist in the entire range \m{\theta \in (0, \upi/2)}. }
\end{figure}

As discussed in \Sec{sec:truncb}, naive \m{Q}MTs assume the perfectly reflecting boundary conditions \eq{eq:bc} for spectral waves, thus precluding PSWs and ignoring potential dissipation at large \m{|n|}. One can expect that, instead of \Eq{eq:bc}, it would be more accurate to adopt the following closure in anticipation of spectral waves:
\begin{gather}
 \amp_{N+1}^\sigma = \ee^{\ii K^\sigma} \amp_N^\sigma
\end{gather}
(and similarly for \m{\amp_{-(N+1)}^\sigma}). The value of \m{K^\sigma} is found, for given \m{\omega}, from \Eq{eq:disprel}:
\begin{gather}
 \ee^{\ii K^\sigma} = \ee^{-\ii\sigma\theta}
 \Big[\ii\Omega^\sigma + \sqrt{1 - (\Omega^\sigma)^2}\Big],
\end{gather}
where \m{\Omega^\sigma\doteq \omega/\omega_0^\sigma}, and we restrict our attention to the range \m{|\re\Omega\,|\leq1} to avoid discontinuities, a choice which is consistent with the range of observed solutions. The sign of the square root is chosen to enforce outward propagation at \m{\im\omega=0}, which also enforces that unstable solutions exponentially decay in the direction of propagation \m{|\exp(\ii K^\sigma_+)|<1}.\footnote{A different choice of signs would also yield formally valid solutions, but would not have the properties that correspond to PSWs of physical interest.}
(Conversely, damped solutions grow in the direction of propagation.) As seen from \Fig{fig:swclosure}, the resulting closure
\begin{gather}\label{eq:swclosure}
 \amp_{N+1}^\sigma = \ee^{-\ii\sigma\theta}\,
 \Big[\ii\Omega^\sigma + \sqrt{1 - (\Omega^\sigma)^2}\Big]
 \amp_N^\sigma
\end{gather}
leads to reasonably accurate results already \m{N = 2}, both for MIs and PSWs. 

Because the closure becomes exact only in the limit \m{N \rightarrow \infty}, it also yields spurious solutions (gray curves in \Fig{fig:swclosure}) at finite \m{N}. One can understand this from the analogy with the Case--van~Kampen modes mentioned in \Sec{sec:global}. These solutions become increasingly stable, and thus less of an issue, as \m{N} increases. Also note that the closure adequately describes true MIs. This is explained by the fact that \m{\amp_n^\sigma} at large \m{|n|} are exponentially small and do not matter for unstable modes anyway.

\section{Quasilinear approximation beyond ideal MHD}
\label{sec:dispersion}

As discussed in the previous sections, the existence of PSWs undermines the standard quasilinear approximation in application to ideal incompressible MHD. Given the ubiquity of quasilinear modeling in the literature, it may seem concerning that the quasilinear approximation can fail so spectacularly. Interestingly, though, the quasilinear approximation is somewhat more robust beyond the ideal-MHD limit, both due to conservative corrections and dissipation.

Let us discuss the former first. Without attempting to describe any particular physical system, let us consider the following modified version of \Eq{eq:elsasser}:
\begin{gather}\label{eq:withdisp}
 \pd_t \vec{z}^\pm = -(\vec{z}^\mp\cdot\nabla)\vec{z}^\pm - \nabla P + \lambda \pd_x\nabla^2\vec{z}^\pm,
\end{gather}
where \m{\lambda} is a constant parameter. The last term is intended as a simple ad~hoc correction that, while causing deviation from ideal MHD, leaves the primary-mode evolution unaffected and preserves MHD's key invariants, specifically, the energy and cross helicity (\App{app:lambda}). Perhaps notably, this term introduces rotational asymmetry in the \m{(x, y)} plane. Similar terms can appear due to a background magnetic field or differential rotation \citep{ref:heinonen23}. 

\begin{figure}
 \centering
 \includegraphics{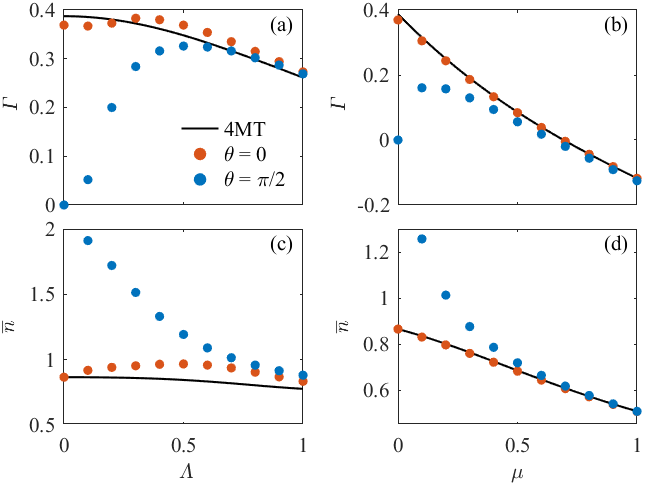}
 \caption{\label{fig:dispscan}
 The growth rate \m{\Gamma}, at \m{\nu_- = 0}, versus: (a)~\m{\Lambda\doteq\lambda/\mc{A}p^3} and (b)~\m{\mu\doteq\nu_+/\mc{A}p^2}. The corresponding `center of energy" of the unstable eigenmode, \m{\overbar{n}\doteq \sqrt{\sum_n \mc{E}_n n^2/\mc{E}}}, is shown in~(c) and~(d), respectively. In all figures, the colour markers indicate the results obtained through DNS of \Eq{eq:matrixeq}, while the black solid curves indicate solutions obtained from the 4MT truncation of \Eq{eq:intermediatewithdisp}. The results are presented for the representative cases \m{\theta = 0} and \m{\theta = \upi/2}, both at \m{r = 0.5} and \m{\phi = \upi/4}.
 }
\end{figure}

Taking \m{\phi = \upi/4} for simplicity, one arrives at the following corresponding modification of the linear \Eq{eq:intermediate}:
\begin{gather}\label{eq:intermediatewithdisp}
    \sqrt{2}(\pd_\tau+ \ii \delta_n)\amp_n^\pm=\quad \ee^{\mp\ii\theta} \alpha_n^- \amp_{n-1}^\pm +\ee^{\pm \ii\theta}\beta_n^- \amp_{n-1}^\mp 
    + \ee^{\pm\ii\theta}\alpha_n^+ \amp_{n+1}^\pm +\ee^{\mp \ii\theta}\beta_n^+ \amp_{n+1}^\mp,
\end{gather}
where \m{\delta_n^\pm \doteq \Lambda r(r^2+n^2)}, \m{\Lambda\doteq\lambda/\mc{A}p^3}, and \m{\alpha_n^\pm}, \m{\beta_n^\pm} are as in \eq{eq:ab}. In the large-\m{|n|} limit, one has \m{\alpha_n^ \pm \sim 1}, \m{\beta_n^\pm \to 0}, \m{\delta_n \sim \Lambda n^2}, so one obtains the following scaling:
\begin{gather}\label{eq:harmonicsuppression}
 |\amp_{n+1}^\pm| \sim \frac{|\amp_n^\pm|}{\Lambda n^2}.
\end{gather}
This shows that the harmonic magnitude \m{|\amp_n|} decreases rapidly (super-exponentially) with \m{|n|}, and thus low-order truncations may be justified.  Note that although the opposite scaling, \m{|\amp_{n-1}^\pm| \sim |\amp_n^\pm|/\Lambda n^2} is also formally possible, such modes cannot be excited, as is the case with inward propagating PSWs. Figure~\ref{fig:dispscan}(a) shows that, indeed, the agreement between analytical QL growth rates and nonlinear DNS improves as the parameter \m{\Lambda} increases. The agreement becomes nearly perfect for \m{\Lambda \gtrsim 0.5}.  Figure~\ref{fig:dispscan}(b) shows that the same effect can be achieved if, instead of the \m{\lambda} term in \Eq{eq:withdisp}, one introduces sufficiently strong viscosity. In this case, one also has a modified \Eq{eq:intermediate} with the exact form of \Eq{eq:intermediatewithdisp}, but with \m{\delta_n^\pm \doteq \mu (r^2+n^2)}, where \m{\mu\doteq\nu_+/\mc{A}p^2} (with \m{\nu_-=0} for simplicity). Again, the agreement becomes nearly exact for \m{\mu \gtrsim 0.5}.

\section{Summary}\label{sec:summary}

In this paper, we explore structure formation in two-dimensional MHD turbulence as a modulational instability (MI) of turbulent fluctuations. We focus on the early stages of structure formation and consider simple backgrounds that allow for a tractable model of the MI while retaining the full chain of modulational harmonics. This approach allows for a systematic examination of the importance of high-order correlations that are typically ignored in mean-field theories. 

We find that, when the primary structure truly experiences a MI, this MI can be described well with typical truncated models that neglect high-order correlations. However, already in adjacent regimes, such truncated models can fail spectacularly and produce false positives for instability. To study this process in detail, we propose an `extended' quasilinear theory (XQLT) that treats the primary structure as fixed but includes the entire spectrum of modulational harmonics (as opposed to just the low-order harmonics, as usual). From XQLT, also corroborated by DNS, we find that the difference between said regimes is due to a fundamental difference in the structures of the modulational spectra. For unstable modes, the spectrum is exponentially localised at low harmonic numbers, so truncated models are justified. But this localization does not always occur. At other parameters, modulational modes turn into constant-amplitude waves propagating down the spectrum, unimpeded until dissipative scales. These spectral waves are self-maintained as global modes with real frequencies and cause ballistic energy transport along the spectrum, breaking the feedback loops that could otherwise sustain MI. 

The ballistic transport by PSWs drains energy from the primary structure at a constant rate until the primary structure is depleted. Because global PSWs exist at almost all parameters, this means that almost any primary structure in ideal incompressible MHD will eventually be depleted. This means, in particular, that sustainability of MHD structures is not entirely limited to the issue of exponentially growing linear instabilities, as PSWs must also be taken into consideration. To describe them within a reduced model, we propose a closure that successfully captures both the propagating spectral waves and~MIs for the assumed primary structure.

Finally, we find that departures from ideal MHD constrains the form of modulational eigenmodes, in turn suppressing the amplitude and impact of high harmonics. This allows us to end on an informed yet optimistic note regarding the applicability of the quasilinear approximation for structure formation in dispersive forms of MHD. That said, it is an important conclusion of our work that, unless deviations from ideal incompressible MHD are substantial, changing the turbulence parameters even slightly can utterly destroy the applicability of an otherwise workable reduced model. An understanding of the complex modulational dynamics supported by (nearly) incompressible MHD provides important context in the interpretation of existing simple closures and potentially opens the path to building more reliable alternatives.

This research was supported by the U.S.\ Department of Energy through contract No.\ DE-AC02-09CH11466.

\appendix

\section{Quasilinear time scale}
\label{app:timescales}

The quasilinear approximation holds when the quasilinear MI growth rate exceeds the rate at which energy escapes to higher harmonics. In the main text, we take the latter to be the group velocity of the PSW global mode. However, one might object that the properties of PSWs are well defined only at large \m{n}, while the departure from quasilinear occurs at the modest \m{n = 2}. It may be instructive then to develop an alternative argument that would not assume large \m{n}. Here, we propose such an argument by considering the initial-value problem.

For simplicity, let us assume the initial conditions such that only \m{\vec{\psi}_0(\tau = 0)} is nonzero, with all other \m{\vec{\psi}_{n\neq 0}(\tau = 0)} are zero. We also adopt \m{(\phi, \theta) = (\upi/4, 0)} and \m{(\phi, \theta) = (\upi/4, \upi/2)} as representative cases for VDPMs and BDPMs respectively. For these parameters, \Eq{eq:general} can be further reduced due to the parity of the initial conditions. Also, the coupling matrices \m{\matr{G_n^\sigma}} become particularly simple:
\begin{gather}
 \matr{G_n^\sigma}=
 \begin{cases}
  \frac{1}{\sqrt{2}}
  \begin{pmatrix}
   \alpha_n^\sigma & \beta_n^\sigma\\
   \beta_n^\sigma &\alpha_n^\sigma
  \end{pmatrix},
  &
  \theta = 0,
  \\
  \frac{\ii \sigma}{\sqrt{2}}
  \begin{pmatrix}
   -\alpha_n^\sigma & \beta_n^\sigma\\
   -\beta_n^\sigma & \alpha_n^\sigma
  \end{pmatrix},
  &
  \theta = \frac{\upi}{2}.
 \end{cases}
\end{gather}

For the case \m{\theta = 0}, we assume assume \m{\psi_0^+ = \psi_0^-}  at \m{\tau = 0}. Then, \m{\dot{\psi}_n^+(\tau = 0) = \dot{\psi}_n^-(\tau = 0)} as well, and thus \m{\psi_n^+ = \psi_n^-} at all times, \ie \m{\vec{b} \equiv 0}. We will refer to this case as \m{\vec{v}}-only initial conditions (VIC). The corresponding dynamics can be described by a single function \m{\psi_n \doteq \psi_n^+ = \psi_n^-}, which satisfies the equation
\begin{gather}
 \dot\psi_n = 
 \frac{\alpha_n^- + \beta_n^-}{\sqrt{2}}\,\psi_{n-1}
 + \frac{\alpha_n^+ + \beta_n^+}{\sqrt{2}}\,\psi_{n+1}.
 \label{eq:psiVIC}
\end{gather}
For the case \m{\theta = \upi/2}, we assume \m{\psi_0^+ = -\psi_0^-} at \m{\tau = 0}. Similarly to the VIC case, this implies that \m{\vec{v} \equiv 0}. We will refer to this case as \m{\vec{b}}-only initial conditions (BIC). The corresponding dynamics can be described by a single function \m{\psi_n \doteq - \ii^n\psi_n^+ = (-\ii)^n \psi_n^-}, which satisfies the equation
\begin{gather}
 \dot\psi_n = 
 \frac{\alpha_n^- - (-1)^n\beta_n^-}{\sqrt{2}}\,\psi_{n-1}
 + \frac{\alpha_n^+ - (-1)^n\beta_n^+}{\sqrt{2}}\,\psi_{n+1}.
 \label{eq:psiBIC}
\end{gather}

To estimate the nonlinear time scale \m{\tau_{\text{NL}}} on which \m{\psi_2} might become comparable with \m{\psi_0}, let us consider \Eqs{eq:psiVIC} and \eq{eq:psiBIC} for \m{n = 2} at early times, when \m{\psi_3} remains negligible:
\begin{gather}
 \ddot{\psi_2} = 
 \begin{cases}
  \frac{\alpha_2^- + \beta_2^-}{\sqrt{2}}
  \left(
  \frac{\alpha_1^- + \beta_1^-}{\sqrt{2}}\,\psi_0
  + \frac{\alpha_1^+ + \beta_1^+}{\sqrt{2}}\,\psi_2
  \right)
  & 
  \text{for VIC},
  \\
  \frac{\alpha_2^- - \beta_2^-}{\sqrt{2}}
  \left(
  \frac{\alpha_1^- + \beta_1^-}{\sqrt{2}}\,\psi_0
  + \frac{\alpha_1^+ + \beta_1^+}{\sqrt{2}}\,\psi_2
  \right)
  &
  \text{for BIC}.
 \end{cases}
\end{gather}
Initially, \m{|\psi_0| \gg |\psi_2|}, so the second term in the brackets can be neglected in both cases and the nonlinear time scale can be readily estimated:
\begin{gather}
 \tau_{\text{NL}}^{-2} = \frac{1}{2}\times
 \begin{cases}
  (\alpha_2^- + \beta_2^-)(\alpha_1^- + \beta_1^-)
  &
  \text{for VIC},
  \\
  (\alpha_2^- - \beta_2^-)(\alpha_1^- + \beta_1^-)
  &
  \text{for BIC}.
 \end{cases}
\end{gather}
In contrast, the quasilinear time scale \m{\tau_{\text{QL}}}, which follows from the 4MT equations \eq{eq:4mt}, is identical in both cases:
\begin{gather}
 \tau_{\text{QL}}^{-2} = \alpha_0^-(\alpha_1^- + \beta_1^-).
\end{gather}
The ratio of time scales is then
\begin{gather}
 \left(\frac{\tau_{\text{QL}}}{\tau_{\text{NL}}}\right)^2 =
 \begin{cases}
  2\sqrt{r^2+4} & \text{for VIC},\\
  \frac{2r\sqrt{r^2+4}}{r+4} & \text{for BIC}.
 \end{cases}
\end{gather}
For the range of interest (\m{|r| < 1}), this always yields
\begin{gather}
\frac{\tau_{\text{QL}}}{\tau_{\text{NL,BIC}}}
< 1 <
\frac{\tau_{\text{QL}}}{\tau_{\text{NL,VIC}}}.
\end{gather}
(For example, at \m{r = 0.5}, one has \m{\tau_{\text{QL}}/\tau_{\text{NL,VIC}} \approx 0.5}, and \m{\tau_{\text{QL}}/\tau_{\text{NL,BIC}} \approx 1.5}.) Thus, quasilinear is sufficient for VIC but not for BIC, in agreement with our numerical results.

\section{Properties of \Eq{eq:withdisp}}
\label{app:lambda}

The operator introduced in \Eq{eq:withdisp} conserves energy (here it is normalised to the constant mass density):
\begin{gather}
 E \doteq \frac{1}{4} \int\dd\vec{x}\,(|\vec{z}^+|^2+|\vec{z}^-|^2),
\end{gather}
and cross helicity:
\begin{gather}
 H_C \doteq \frac{1}{4} \int\dd\vec{x}\,(|\vec{z}^+|^2-|\vec{z}^-|^2).
\end{gather}
These can be expressed in terms of the Elsasser energies
\begin{gather}
 E^\pm \doteq \int\dd\vec{x}\,|\vec{z}^\pm|^2/4,
\end{gather}
as
\begin{gather}
 E=E^++E^-, \qquad H_C=E^+-E^-.
\end{gather}
Equation \eq{eq:withdisp} yields
\begin{align}
\begin{split}
 \frac{\dd E^\pm}{\dd t}
 & = \frac{1}{4}\int\dd\vec{x}\,\pd_t |\vec{z}^\pm|^2\\
 & = \frac{1}{2}\int\dd\vec{x}\,\vec{z}^\pm\cdot\pd_t\vec{z}^\pm\\
 & = \frac{\lambda}{2}\int\dd\vec{x}\,\vec{z}^\pm\cdot\pd_x\nabla^2\vec{z}^\pm\\
 & = \frac{\lambda}{2}\int\dd\vec{x}\,z^\pm_l\pd_x\pd_m^2z^\pm_l\\
 & = -\frac{\lambda}{2}\int\dd\vec{x}\,(\pd_m z^\pm_l) (\pd_x\pd_mz^\pm_l)\\
 & = -\frac{\lambda}{4}\int\dd\vec{x}\pd_x\,(\pd_m z^\pm_l\pd_m z^\pm_l)\\
 & = 0.
\end{split}
\end{align}
Thus,
\begin{gather}
 \frac{\dd E}{\dd t} =  \frac{\dd E^+}{\dd t} +  \frac{\dd E^-}{\dd t} = 0,\\
 \frac{\dd H_C}{\dd t} =  \frac{\dd E^+}{\dd t} -  \frac{\dd E^-}{\dd t} = 0,
\end{gather}
meaning that \Eq{eq:withdisp} conserves both \m{E} and \m{H_C}.

\bibliographystyle{jpp}

\bibliography{bib-mhd,bib-my,bib-main}

\end{document}